\documentclass[prb,twocolumn,eqsecnum,showpacs,floatfix,amsmath,superscriptaddress]{revtex4}

\usepackage{bm} 
\usepackage{graphicx} 
\usepackage[dvipdfm,letterpaper,textwidth=7.0in,textheight=9.5in]{geometry}
\usepackage{hyperref}

\newcommand{\degree}{^{\circ}}

\begin{document}


\title{Phase diagram of the three-band half-filled Cu-O two-leg ladder}


\date{\today}
\author{Sootaek Lee}
\author{J. B. Marston}
\affiliation{Department of Physics, Brown University, Providence, Rhode Island 02912-1843, USA}
\author{J. O. Fj{\ae}restad}
\affiliation{Department of Physics and Astronomy, University of California, Los Angeles, California 90095, USA}
\affiliation{Department of Physics, University of Queensland, Brisbane, Qld 4072, Australia}


\begin{abstract}
We determine the phase diagram of the half-filled two-leg ladder both at weak and strong coupling, taking into account the Cu $d_{x^2-y^2}$ and the O $p_x$ and $p_y$ orbitals.  At weak coupling, renormalization group flows are interpreted with the use of bosonization.  Two different models with and without outer oxygen orbitals are examined. For physical parameters, and in the absence of the outer oxygen orbitals, the D-Mott phase arises; a dimerized phase appears when the outer oxygen atoms are included. We show that the circulating current phase that preserves translational symmetry does not appear at weak coupling. In the opposite strong-coupling atomic limit the model is purely electrostatic and the ground states may be found by simple energy minimization. The phase diagram so obtained is compared to the weak-coupling one.
\end{abstract}


\maketitle


\section{Introduction}

Considerable effort has been expended to explain the phase diagram of the high temperature superconductors and the associated pseudogap phenomenon within the framework of quantum critical and competing order pictures. Among the several candidates for orders that could compete with superconductivity are the ``circulating current (CC) phase,''\cite{varma97a, varma99} the ``staggered flux (SF) phase''\cite{affleck88, marston89, hsu91, wen96} or ``$d$-density wave (DDW) phase,''\cite{nayak00, chakravarty01} the ``spin-Peierls phase,''\cite{vojta99, park01} ``stripes''\cite{emery97, zaanen99} or the ``quantum liquid crystal phase,''\cite{kivelson98, zhang02} and the antiferromagnetic phase.\cite{zhang97a}  The CC phase, like the SF phase, breaks time reversal symmetry and is characterized by circulating currents which produce local orbital magnetic moments. An important difference between the two, however, is that the CC phase preserves translational symmetry while the SF phase breaks it.

It is a nontrivial task to ascertain phase diagrams of strongly correlated systems in two dimensions due to the lack of accurate, systematic, nonperturbative methods. Mean-field type approximations always favor from the outset particular types of orders. Exact diagonalization is constrained by small system size. Quantum Monte-Carlo methods have the notorious fermion sign problem. On the other hand, there are reliable nonperturbative methods available in one dimension such as bosonization\cite{delft98, gogolin98, giamarchi04} and the density matrix renormalization group (DMRG) method.\cite{white93, peschel98, schollwock04} By studying ladder systems one can take a first step toward investigating some of the theoretical ideas that have been conjectured for two-dimensional systems, while remaining in an essentially one-dimensional setting. Moreover, studies of ladder systems have been strongly stimulated by synthetic compounds with the ladder structure. For example, Sr$_{14}$Cu$_{24}$O$_{41}$ exhibits superconductivity upon Ca doping and the application of pressure.\cite{uehara96}

Soon after the discovery of high-temperature superconductivity, one-band models such as the Hubbard or $t$-$J$ model were suggested as the simplest microscopic systems that could capture the interesting physics of the Cu-O system.\cite{anderson87}  Indeed, most work on ladders has focused on one-band models that consider only the Cu $d$ orbital.\cite{dagotto96, schulz96, balents96, lin98, orignac97, fjaerestad02, tsuchiizu02, wu03, white02, marston03, schollwock03, fjaerestad04, momoi03} However, there are a number of theoretical studies\cite{zhang97b, pryadko03} that suggest that neither the two-dimensional Hubbard nor $t$-$J$ models have a superconducting ground state.  It is also impossible to capture some other types of potential order like the CC phase within a one-band model as the circulating current pattern in the CC phase involves the O orbitals in a crucial way.\cite{srinivasan02}  A more complete model that incorporates these orbitals is the three band or Emery model.\cite{emery87, varma87, varma97a} The three-band model takes into account both strong correlations and the hybridization of Cu and O orbitals.  

Relatively little work has been done on three-band ladder models so far.\cite{srinivasan02, nishimoto02} In this paper we obtain phase diagrams of the half-filled Cu-O ladder both at weak coupling and at strong coupling. At weak coupling, we examine ladder systems of two different geometries: with and without outer oxygen orbitals. Renormalization group (RG) flows are interpreted with the use of bosonization followed by semiclassical energy minimization. Several gapped phases arise depending on the particular values of the parameters. Within the physically relevant region, the D-Mott phase, which upon doping has a tendency toward $d$-wave superconductivity, arises in the absence of the outer oxygen sites while a spin Peierls phase occurs when outer oxygens are added. We find that the CC phase does not appear at weak coupling in the Cu-O ladder system.  At strong coupling, the model is purely electrostatic. Treating the hopping term as a weak perturbation, it is then possible to make connections between the weak and strong coupling phase diagrams.

The paper is organized as follows. In Sec.~\ref{sec:model}, we define the models and discuss the weak-coupling continuum limit. We present the RG flow equations in Sec.~\ref{sec:rg}. In Sec.~\ref{sec:bosonization}, we bosonize the Hamiltonian and the various local order parameters of interest. From these, in Sec.~\ref{sec:weak} the phase diagrams of the Cu-O ladders are established at weak coupling. The volumes of the different phases in the high-dimensional parameter space is quantified by means of Monte-Carlo sampling. In Sec.~\ref{sec:strong}, we present the phase diagram at strong coupling and compare our findings to those at weak coupling. Our results are summarized in concluding Sec.~\ref{sec:conclusion}. Some technical details are relegated to the Appendix.


\section{Model}\label{sec:model}

\begin{figure}
\includegraphics[width=3.0in]{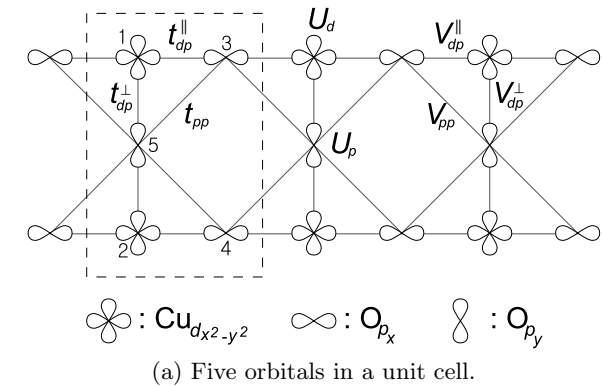}\\
(a) Five orbitals in a unit cell.\\
$\phantom{\frac{1}{1}}$ \\
\includegraphics[width=3.2in]{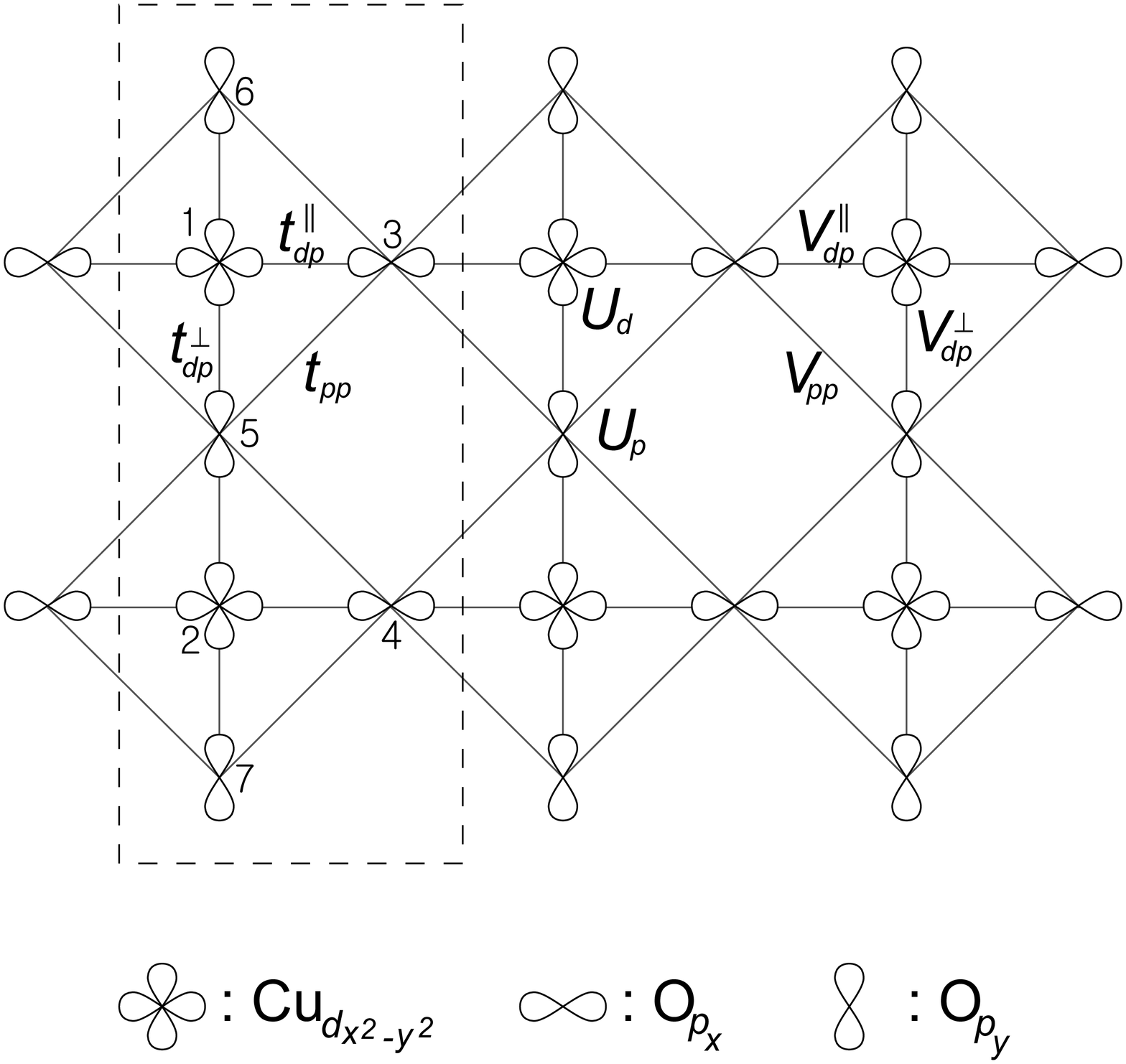}\\
(b) Seven orbitals in a unit cell.\\
\caption{Schematic diagrams of the Cu-O ladder.
(a) The five orbital case: There are two Cu $d$ orbitals and three O $p$ orbitals in a unit cell. Hopping integrals $t_{dp}^{\parallel}$ and $t_{dp}^{\perp}$ are between Cu $d$ and O $p$; $t_{pp}$ is the hopping between the oxygens. $U_d$ and $U_p$ are on-site Coulomb interactions, and $V_{dp}^{\parallel}$, $V_{dp}^{\perp}$ and $V_{pp}$ are nearest-neighbor Coulomb interactions.
(b) The seven orbital case: There are two extra outer oxygen orbitals in a unit cell labeled 6 and 7.}
\label{fig:ladder}
\end{figure}

We study half-filled Cu-O two-leg ladders with Cu $d_{x^2-y^2}$ and O $p_x$ and $p_y$ orbitals as shown in Fig.~\ref{fig:ladder}. We consider two different geometries: (1) the five orbital case with two Cu $d_{x^2-y^2}$ orbitals ($i=1,2$), two O $p_{x}$ orbitals ($i=3,4$), and one O $p_y$ orbital ($i=5$) in a unit cell [see Fig.~\ref{fig:ladder}(a)] and (2) the seven orbital case in which in addition to the five orbitals there are also two outer oxygen $p_y$ orbitals ($i=6,7$) [see Fig.~\ref{fig:ladder}(b)]. At half-filling, on the average, there are eight electrons in a unit cell in the five orbital model and twelve electrons in the seven orbital model.

The Hamiltonian we investigate in this paper is given by
\begin{equation}\label{eq:hamiltonian}
H=H_0+H_{I},
\end{equation}
where
{\setlength\arraycolsep{2pt}
\begin{eqnarray}\label{eq:h0}
H_0 
&=& \sum_{x,i} \epsilon_i n_i(x) 
- \sum_{x,i,j,\sigma} 
\big[ t^{intra}_{ij} c^{\dagger}_{i\sigma}(x) c^{}_{j\sigma}(x) \nonumber \\
& & + t^{inter}_{ij} c^{\dagger}_{i\sigma}(x) c^{}_{j\sigma}(x-1) + \textrm{h.c.} \big]
\end{eqnarray}}
is the tight-binding Hamiltonian and
{\setlength\arraycolsep{2pt}
\begin{eqnarray}\label{eq:hi}
H_{I} 
&=& \sum_{x,i} \frac{U_{i}}{2} n_{i}(x) \left[ n_{i}(x) - 1 \right]
+ \sum_{x,i,j} \big[ V^{intra}_{ij} n_i(x) n_j(x) 
\nonumber \\
&& + V^{inter}_{ij} n_i(x) n_j(x-1) \big]
\end{eqnarray}}
is the Coulomb interaction. Each site is labeled by two integers $(i,x)$ where ``$i$'' labels atoms within a unit cell, and ``$x$'' identifies different cells. Operators $c^{}_{i\sigma}(x)$ and $c^{\dagger}_{i\sigma}(x)$, respectively, annihilate and create either Cu $d$ electrons ($i=1, 2$) or O $p$ electrons ($i=3, 4, 5$ in the five orbital case and $i=3, 4, 5, 6, 7$ in the seven orbital case) in $x$th unit cell with spin $\sigma$ ($\sigma=\uparrow, \downarrow$). The number operator $n_i(x)=\sum_{\sigma}c^{\dagger}_{i\sigma}(x)c^{}_{i\sigma}(x)$ counts electrons at site ($i, x$).

Consider now the hopping amplitudes between neighboring Cu and O sites, $t^{\parallel}_{dp}$ and $t^{\perp}_{dp}$, as well as between nearest O and O sites, $t^{}_{pp}$. Since there is no particle-hole symmetry, the sign of each hopping term is relevant. Signs of the various hopping parameters are determined by the symmetry of the orbitals. By choosing appropriate phases for each orbital, the hopping parameters $t^{intra}_{ij}$ and $t^{inter}_{ij}$ in Eq.~(\ref{eq:h0}) for the five orbital case may be all be taken to be positive
\begin{subequations}
\label{eq:hopping}
\begin{equation} \label{eq:t_intra}
t^{intra}_{ij} = \left\{ \begin{array}{ll}
t^{\parallel}_{dp} & ~\textrm{if}~ (i,j) = (1,3)~\textrm{or}~(2,4) \\
t^{\perp}_{dp} & ~\textrm{if}~ (i,j) = (1,5)~\textrm{or}~(2,5) \\
t^{}_{pp} & ~\textrm{if}~ (i,j) = (3,5)~\textrm{or}~(4,5) \\
0 & ~\textrm{otherwise,}
\end{array} \right.
\end{equation}
\begin{equation} \label{eq:t_inter}
t^{inter}_{ij} = \left\{ \begin{array}{ll}
t^{\parallel}_{dp} & ~\textrm{if}~ (i,j) = (1,3)~\textrm{or}~(2,4) \\
t^{}_{pp} & ~\textrm{if}~ (i,j) = (3,5)~\textrm{or}~(4,5) \\
0 & ~\textrm{otherwise.}
\end{array} \right.
\end{equation}
\end{subequations}
Hopping parameters in the seven orbital case follow a similar pattern. The on-site energy of an electron on a Cu or O site is given by $\epsilon_d$ or $\epsilon_p$ respectively. For guidance we choose on-site energies and hopping matrix elements as extracted from a density functional theory (DFT) calculation for YBCO by Andersen {\em et al.}:\cite{okandersen95} $\epsilon = \epsilon_d - \epsilon_p = 3.0$ eV, $t_{dp} = 1.6$ eV, and $t_{pp} = 1.1$ eV. Although the precise values of these parameters for real ladder compounds will differ from those of the full two-dimensional problem studied by Andersen {\em et al.}, for concreteness we use the same values here for the ladder. Thus for example the hopping along the legs $t^{\parallel}_{dp}$ and along the rungs $t^{\perp}_{dp}$ are taken to be the same, though this is not required by symmetry.

Turning now to the Coulomb interactions, the on-site energies $U_i$ take two different values, one for Cu atoms, the other for the O atoms 
\begin{equation} \label{eq:ui}
U_i= \left\{ \begin{array}{ll}
U_d & ~\textrm{if}~ i=1,2 \\
U_p & ~\textrm{if}~ i=3,4,5.
\end{array} \right.
\end{equation}
Nearest-neighbor interactions are
\begin{subequations}
\label{eq:nnv}
\begin{equation} \label{eq:v_intra}
V^{intra}_{ij} = \left\{ \begin{array}{ll}
V^{\parallel}_{dp} & ~\textrm{if}~ (i,j) = (1,3)~\textrm{or}~(2,4) \\
V^{\perp}_{dp} & ~\textrm{if}~ (i,j) = (1,5)~\textrm{or}~(2,5) \\
V^{}_{pp} & ~\textrm{if}~ (i,j) = (3,5)~\textrm{or}~(4,5) \\
0 & ~\textrm{otherwise,}
\end{array} \right.
\end{equation}
\begin{equation} \label{eq:v_inter}
V^{inter}_{ij} = \left\{ \begin{array}{ll}
V^{\parallel}_{dp} & ~\textrm{if}~ (i,j) = (1,3)~\textrm{or}~(2,4) \\
V^{}_{pp} & ~\textrm{if}~ (i,j) = (3,5)~\textrm{or}~(4,5) \\
0 & ~\textrm{otherwise.}
\end{array} \right.
\end{equation}
\end{subequations}
The case of seven orbitals again follows a similar pattern.

The Hamiltonian has the usual U(1)$\times$SU(2) global charge/spin symmetry. Furthermore, it is invariant under lattice translations, time reversal, parity, and chain interchange operations. As stated above, there is no particle-hole symmetry in this system in contrast to the one-band model. Since the symmetry differs, the RG equations also differ, as discussed below in Sec.~\ref{sec:rg}.


\subsection{Band Structure}

\begin{figure}
\includegraphics[width=3.0in]{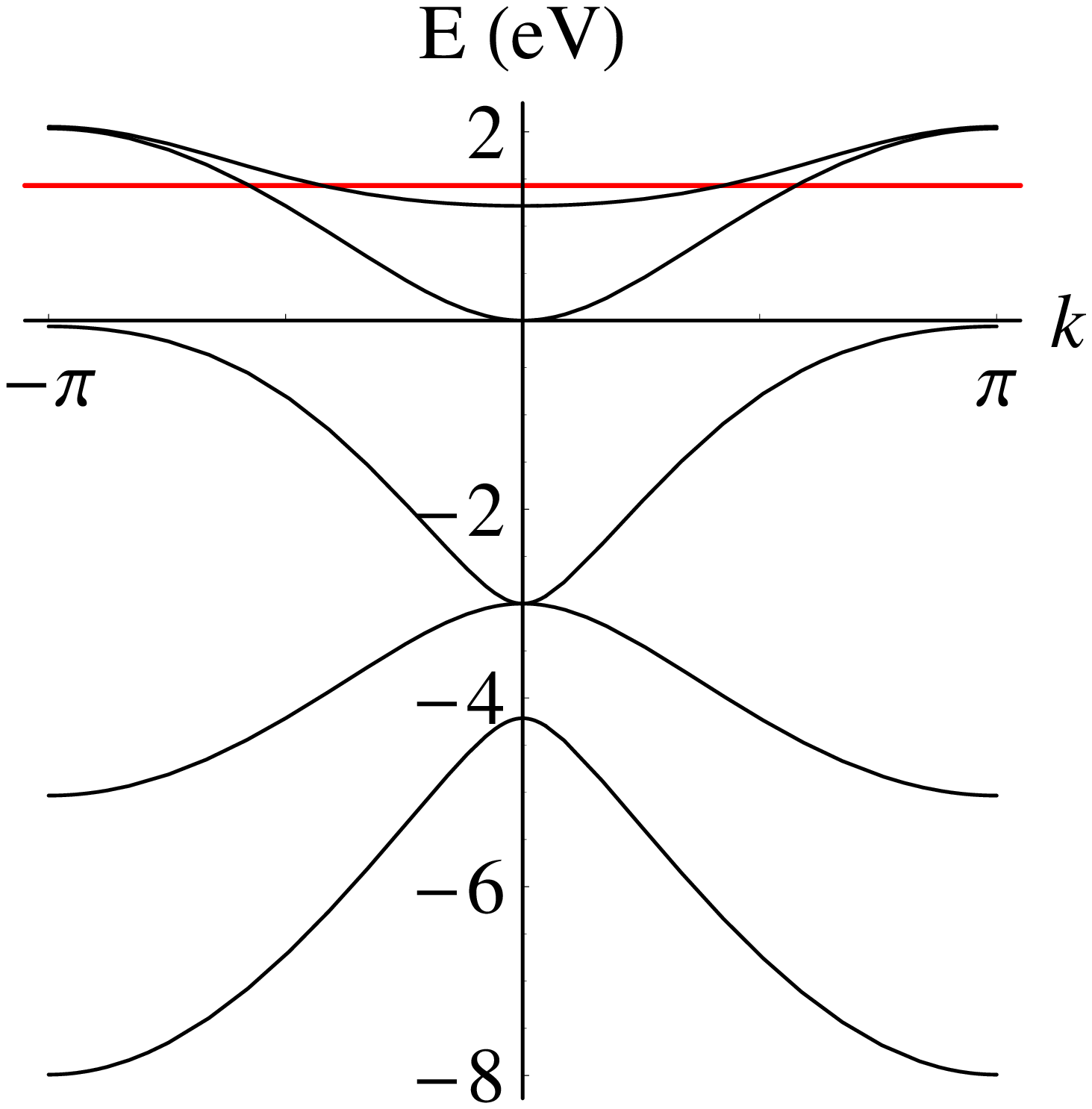}\\
(a) Five orbital case \\
\includegraphics[width=3.0in]{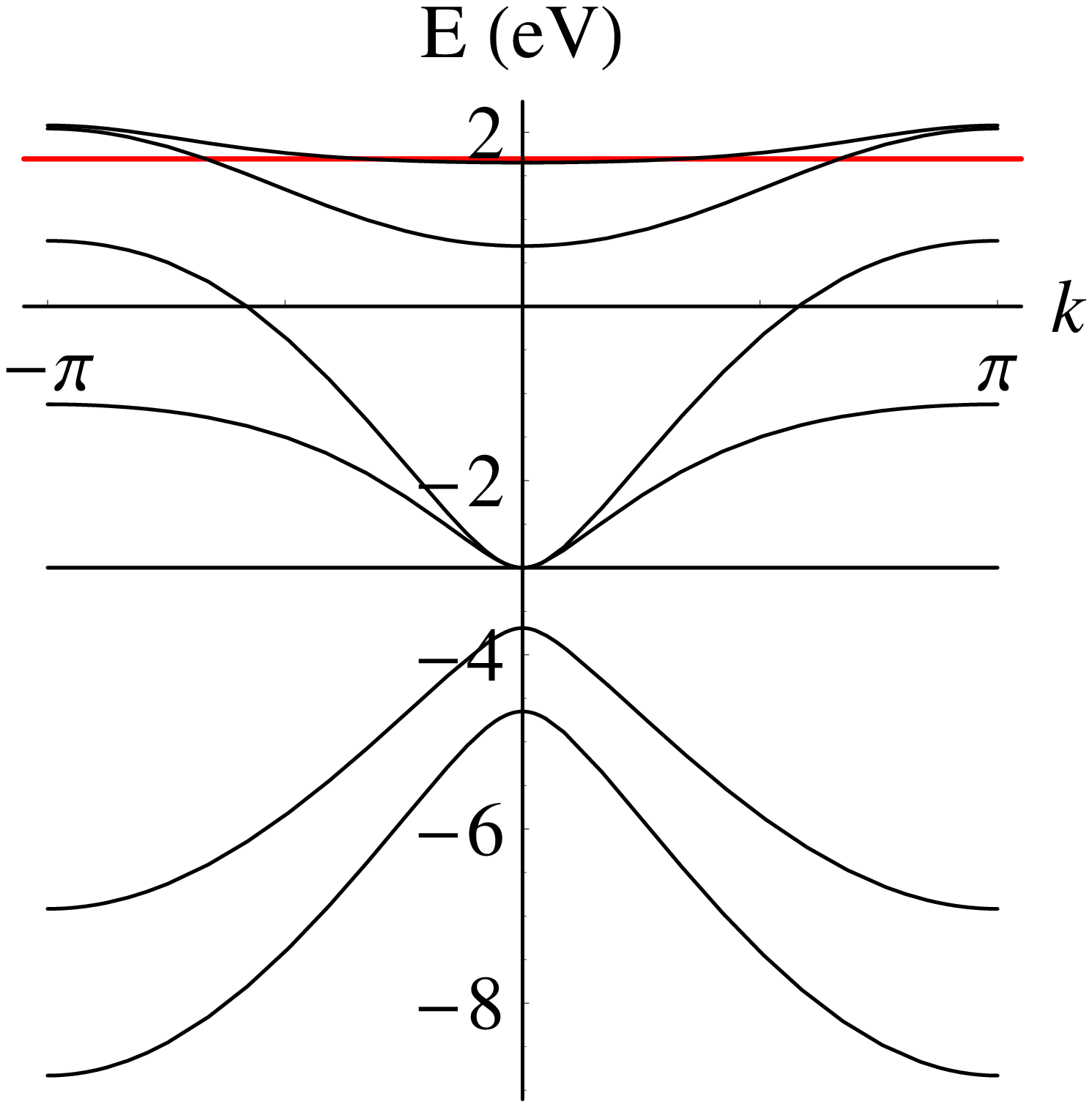}\\
(b) Seven orbital case
\caption{Energy spectrum of the quadratic portion of the Hamiltonian in momentum space.
(a) Five orbital case: Three low-lying bands are completely filled and only the upper two bands intersect the Fermi level. The Fermi velocities at $k_{F1}$ and $k_{F2}$ differ even at half filling unlike the one band model. We use $\epsilon^{}_d - \epsilon^{}_p = 3.0$ eV, $t^{\parallel}_{dp}=t^{\perp}_{dp}=1.6$ eV, and $t^{}_{pp}=1.1$ eV per the DFT calculation of Andersen {\em et al.} (see Ref.~\onlinecite{okandersen95}).
(b) Seven orbital case for the same parameters as the five orbital case.} 
\label{fig:band}
\end{figure}

We first diagonalize the quadratic, noninteracting portion of the Hamiltonian $H_0$ to obtain the band structure of the Cu-O ladder. The band structure is shown in Fig.~\ref{fig:band}. Evident in Fig.~\ref{fig:band} is the fact that only two bands cross the Fermi level so we can analyze both ladders (with and without extra oxygens) in a similar manner. Note, however, that for very different tight-binding parameters different numbers of bands may intersect the Fermi level. In the two band case, it can be easily shown that $k_{F1}+k_{F2}=\pi$ at half filling, where $k_{Fn} > 0$ is a Fermi momentum defined by $\epsilon_{n}(k_{Fn})=\mu$ and $\mu$ is chemical potential. However, since there is no particle-hole symmetry, the Fermi velocity $v_{Fn}=\partial \epsilon_{n}/ \partial k |_{k=k_{Fn}}$ at the two Fermi points differs even at half filling: $v_{F1} \neq v_{F2}$.

The band operator $\psi^{}_{n\sigma}(x)$ is a linear combination of the $c^{}_{i\sigma}(x)$
\begin{equation}
\psi_{n\sigma}(x)=\sum_{i} a^{}_{in} c^{}_{i\sigma}(x),
\end{equation}
where $n=1,2,...,5,(6,7)$ is a band index and $i=1,2,...,5,(6,7)$ is a site index in a unit cell. It is always possible to choose the matrix element $a^{}_{in}$ to be real since the Hamiltonian is time-reversal invariant.  Furthermore, due to the symmetry under chain exchange, we have the following relations between $a^{}_{in}$:
\begin{equation}
|a^{}_{1n}| = |a^{}_{2n}|, \phantom{0} 
|a^{}_{3n}| = |a^{}_{4n}|, \phantom{0} 
|a^{}_{6n}| = |a^{}_{7n}|.
\end{equation}

At weak coupling it suffices to focus only on the four Fermi points.  Therefore, we will consider only the upper two bands ($n=1,2$) in the subsequent calculation, since the completely filled bands are not active. We linearize the bands around the Fermi points and decompose the band operators into right and left moving fermion fields ($\psi^{}_{nR\sigma}/\psi^{}_{nL\sigma}$):
\begin{equation}
\psi^{}_{n\sigma} 
\propto \psi^{}_{nR\sigma} e^{ik_{Fn}x} + \psi^{}_{nL\sigma} e^{-ik_{Fn}x}.
\end{equation}
With this decomposition $H_0$ reduces to
\begin{equation}
H_0 
= - \sum_{n,\sigma} \int dx v_{Fn} 
(\psi^{\dagger}_{nR\sigma} i \partial_{x} \psi^{}_{nR\sigma} 
- \psi^{\dagger}_{nL\sigma} i \partial_{x} \psi^{}_{nL\sigma}).
\end{equation}


\subsection{Interaction Hamiltonian - Current Algebra}

We now discuss the interactions between the electrons around the four Fermi points. We introduce the usual current operators\cite{balents96,current_algebra}
\begin{displaymath}
J^{}_{nP} = :\psi^{\dagger}_{nP\alpha} \psi^{}_{nP\alpha}:, \phantom{0}
{\bm J}^{}_{nP} = \frac{1}{2} :\psi^{\dagger}_{nP\alpha} {\bm \sigma}^{}_{\alpha\beta} \psi^{}_{nP\beta}:,
\end{displaymath}
\begin{displaymath}
L^{}_{P} = \psi^{\dagger}_{1P\alpha} \psi^{}_{2P\alpha}, \phantom{0}
{\bm L}^{}_{P} = \frac{1}{2} \psi^{\dagger}_{1P\alpha} {\bm \sigma}^{}_{\alpha\beta} \psi^{}_{2P\beta},
\end{displaymath}
\begin{equation}\label{eq:currents}
M^{}_{nP} = \frac{1}{2} \psi^{}_{nP\alpha} \sigma^{y}_{\alpha\beta} \psi^{}_{nP\beta}, \phantom{0}
N_{P\alpha\beta} = \psi^{}_{1P\alpha} \psi^{}_{2P\beta},
\end{equation}
where ${\bm \sigma}$ denotes Pauli matrices, $P$ denotes the chirality ($R$ or $L$), and $:\phantom{0}:$ means normal ordering (vacuum subtraction.) Normal ordering signs are suppressed in the subsequent discussion for notational convenience. We also follow the notation of Balents {\em et al.}\cite{balents96} to permit comparison between the RG equations for the one and three-band ladder systems in the Sec.~\ref{sec:rg}.

Interaction Hamiltonian density ${\mathcal H}^{}_{I}$ can be expressed into three parts in terms of their nature. Let ${\mathcal H}^{}_{I} = {\mathcal H}^{(1)}_{I} + {\mathcal H}^{(2)}_{I} + {\mathcal H}^{(3)}_{I}$. There are eight allowed interactions connecting left and right movers, with Hamiltonian densities ${\mathcal H}^{(1)}_{I}$:
{\setlength\arraycolsep{2pt}
\begin{eqnarray}\label{eq:h1}
-{\mathcal H}^{(1)}_I 
&=& \widetilde{g}_{1\rho} J^{}_{1R} J^{}_{1L} 
+ \widetilde{g}_{2\rho} J^{}_{2R} J^{}_{2L}
+ \widetilde{g}_{x\rho} ( J^{}_{1R} J^{}_{2L}
\nonumber \\
&& + J^{}_{2R} J^{}_{1L} )
+ \widetilde{g}_{1\sigma} {\bm J}^{}_{1R} \cdot {\bm J}^{}_{1L}
+ \widetilde{g}_{2\sigma} {\bm J}^{}_{2R} \cdot {\bm J}^{}_{2L}
\nonumber \\
&& + \widetilde{g}_{x\sigma} ( {\bm J}^{}_{1R} \cdot {\bm J}^{}_{2L} 
+ {\bm J}^{}_{2R} \cdot {\bm J}^{}_{1L} )
+ \widetilde{g}_{t\rho} ( L^{}_{R} L^{}_{L}
\nonumber \\
&&+ L^{\dagger}_{R} L^{\dagger}_{L} ) 
+ \widetilde{g}_{t\sigma} ( {\bm L}^{}_{R} \cdot {\bm L}^{}_{L} 
+ {\bm L}^{\dagger}_{R} \cdot {\bm L}^{\dagger}_{L} ).
\end{eqnarray}}
Six additional interactions are completely chiral
{\setlength\arraycolsep{2pt}
\begin{eqnarray}\label{eq:h2}
-{\mathcal H}^{(2)}_I 
&=& \widetilde{\lambda}_{1\rho} ( J^{2}_{1R} + J^{2}_{1L} ) 
+ \widetilde{\lambda}_{2\rho} ( J^{2}_{2R} + J^{2}_{2L} ) 
\nonumber \\
&&+ \widetilde{\lambda}_{x\rho} ( J^{}_{1R} J^{}_{2R} + J^{}_{1L} J^{}_{2L} ) 
+ \widetilde{\lambda}_{1\sigma} ( {\bm J}^{}_{1R} \cdot {\bm J}^{}_{1R}
\nonumber \\
&&+ {\bm J}^{}_{1L} \cdot {\bm J}^{}_{1L} )
+ \widetilde{\lambda}_{2\sigma} ( {\bm J}^{}_{2R} \cdot {\bm J}^{}_{2R}
+ {\bm J}^{}_{2L} \cdot {\bm J}^{}_{2L} )
\nonumber \\
&&+ \widetilde{\lambda}_{x\sigma} ( {\bm J}^{}_{1R} \cdot {\bm J}^{}_{2R} + {\bm J}^{}_{1L} \cdot {\bm J}^{}_{2L} ).
\end{eqnarray}}
The couplings in Eq.~(\ref{eq:h2}) just renormalize the velocities of the charge and spin modes, and can be neglected in our second order calculation.  Additional $M_{nP}$ and $N_{P\alpha\beta}$ operators must be introduced to describe Umklapp processes
{\setlength\arraycolsep{2pt}
\begin{eqnarray}\label{eq:h3}
-{\mathcal H}^{(3)}_I 
&=& \widetilde{g}_{1u} ( M^{\dagger}_{1R} M^{}_{1L} + M^{\dagger}_{1L} M^{}_{1R} ) 
+ \widetilde{g}_{2u} ( M^{\dagger}_{2R} M^{}_{2L}
\nonumber \\
&&+ M^{\dagger}_{2L} M^{}_{2R} )
+ \widetilde{g}_{xu} ( M^{\dagger}_{1R} M^{}_{2L} + M^{}_{1R} M^{\dagger}_{2L}
\nonumber \\
&&+ M^{\dagger}_{2R} M^{}_{1L} + M^{}_{2R} M^{\dagger}_{1L} )
+ \widetilde{g}_{tu1} ( N^{\dagger}_{R\alpha\beta} N^{}_{L\alpha\beta}
\nonumber \\
&&+ N^{}_{R\alpha\beta} N^{\dagger}_{L\alpha\beta} ) 
+ \widetilde{g}_{tu2} ( N^{\dagger}_{R\alpha\beta} N^{}_{L\beta\alpha}
\nonumber \\
&&+ N^{}_{R\alpha\beta} N^{\dagger}_{L\beta\alpha} )
\end{eqnarray}}
At half-filling the three interband Umklapp terms ($\widetilde{g}_{xu}$, $\widetilde{g}_{tu1}$, $\widetilde{g}_{tu2}$) are nonvanishing. The single-band Umklapp term, $\widetilde{g}_{nu}$, is nonzero only if $k_{Fn} = \pi/2$. The detailed relationships between $\widetilde{g}$ and $U$, $V$ are listed in the Appendix.


\section{Renormalization Group} \label{sec:rg}

We employ the RG approach combined with bosonization followed by semiclassical energy minimization to obtain the phase diagram at weak coupling. Comparison of the method to results of essentially exact DMRG calculations has shown it to be qualitatively reliable in cases where agreement has been checked; see for instance Ref.~\onlinecite{fjaerestad04}.

The current algebra as outlined in Balents {\em et al.}\cite{balents96} may be used to obtain one-loop RG flow equations for coupling constants.  At half-filling there are eleven coupling constants, and the full set of equations is given by
\begin{displaymath}
\dot{g}_{1\rho} 
= \beta \left( g^{2}_{t\rho} + \frac{3}{16} g^{2}_{t\sigma} 
- g^{2}_{tu1} - g_{tu1}g_{tu2} - g^{2}_{tu2} \right),
\end{displaymath}
\begin{displaymath}
\dot{g}_{2\rho} 
= \alpha \left( g^{2}_{t\rho} + \frac{3}{16} g^{2}_{t\sigma} 
- g^{2}_{tu1} - g_{tu1}g_{tu2} - g^{2}_{tu2} \right),
\end{displaymath}
\begin{displaymath}
\dot{g}_{x\rho} 
= -g^{2}_{t\rho} - \frac{3}{16} g^{2}_{t\sigma} -
g^{2}_{tu1} - g_{tu1}g_{tu2} - g^{2}_{tu2} - g^{2}_{xu},
\end{displaymath}
\begin{displaymath}
\dot{g}_{1\sigma} 
= \beta \left( 2 g_{t\rho}g_{t\sigma} - \frac{1}{2} g^{2}_{t\sigma} 
- 4 g^{2}_{tu1} - 4 g_{tu1}g_{tu2} \right) 
- \alpha g^{2}_{1\sigma},
\end{displaymath}
\begin{displaymath}
\dot{g}_{2\sigma} 
= \alpha \left( 2 g_{t\rho}g_{t\sigma} - \frac{1}{2} g^{2}_{t\sigma} 
- 4 g^{2}_{tu1} - 4 g_{tu1}g_{tu2} \right) 
- \beta g^{2}_{2\sigma},
\end{displaymath}
\begin{displaymath}
\dot{g}_{x\sigma} 
= -2 g_{t\rho}g_{t\sigma} - \frac{1}{2} g^{2}_{t\sigma} 
-4 g_{tu1}g_{tu2} - 4 g^{2}_{tu2} - g^{2}_{x\sigma},
\end{displaymath}
\begin{displaymath}
\dot{g}_{t\rho} 
= g_{0\rho}g_{t\rho} + \frac{3}{16} g_{0\sigma} g_{t\sigma} 
- g_{xu} ( g_{tu1} - g_{tu2} ),
\end{displaymath}
{\setlength\arraycolsep{2pt}
\begin{eqnarray}
\dot{g}_{t\sigma} 
&=& g_{0\sigma}g_{t\rho} 
+ \left( g_{0\rho} - \frac{1}{2} g_{0\sigma} - 2 g_{x\sigma} \right) g_{t\sigma}
\nonumber \\
&&+ 4g_{xu} ( g_{tu1} + g_{tu2} ),
\nonumber
\end{eqnarray}}
{\setlength\arraycolsep{2pt}
\begin{eqnarray}
\dot{g}_{xu}
&=& - \left( 2 g_{t\rho} - \frac{3}{2} g_{t\sigma} \right) g_{tu1} 
+ \left( 2 g_{t\rho} + \frac{3}{2} g_{t\sigma} \right) g_{tu2}
\nonumber \\
&&- 4 g_{x\rho}g_{xu},
\nonumber
\end{eqnarray}}
{\setlength\arraycolsep{2pt}
\begin{eqnarray}
\dot{g}_{tu1}
&=& - ( 2 g_{t\rho} - \frac{1}{2} g_{t\sigma} ) g_{xu} 
- g_{tu2} g_{x\sigma} 
- g_{tu1} \Big[ 2 g_{x\rho}
\nonumber \\
&&- \frac{1}{2} g_{x\sigma} + \alpha g_{1\rho}
+ \beta g_{2\rho} 
+ \frac{3}{4}(\alpha g_{1\sigma} + \beta g_{2\sigma} ) \Big],
\nonumber
\end{eqnarray}}
{\setlength\arraycolsep{2pt}
\begin{eqnarray}
\dot{g}_{tu2} 
&=& ( 2 g_{t\rho} + g_{t\sigma}/2 ) g_{xu} - g_{tu1}g_{x\sigma} 
- g_{tu2} \Big[ 2 g_{x\rho}
\nonumber \\
&&+ \frac{3}{2} g_{x\sigma} + \alpha g_{1\rho}
+ \beta g_{2\rho}
- \frac{1}{4}(\alpha g_{1\sigma} + \beta g_{2\sigma} ) \Big],
\nonumber \\
&& \label{eq:rgflow}
\end{eqnarray}}
where $g_{i} \equiv \widetilde{g}_{i} / [\pi(v_{F1}+v_{F2})]$, $\alpha \equiv (v_{F1}+v_{F2})/(2 v_{F1})$, $\beta \equiv(v_{F1}+v_{F2})/(2 v_{F2})$, $g_{0\rho} = \alpha g_{1\rho} + \beta g_{2\rho} - 2 g_{x\rho}$, and $g_{0\sigma} = \alpha g_{1\sigma} + \beta g_{2\sigma} - 2 g_{x\sigma}$.  The dot indicates logarithmic derivative with respect to the length scale, {\em i.e.}, $\dot{g}_{i} \equiv \partial g_{i} / \partial s$, where $s = \ln l$.

The set of RG flow equations obtained in this system is different from those obtained in one band case, due to the absence of particle-hole symmetry in Cu-O ladder. Equation (\ref{eq:rgflow}) is the most general form of RG flow equations for half-filled ladder systems with four Fermi points and with parity, time-reversal symmetry, chain interchange, and $U(1) \times SU(2)$ global charge/spin symmetry. Note that Eq.~(\ref{eq:rgflow}) is invariant under band interchange. Equation (\ref{eq:rgflow}) also correctly reproduce the RG equations in the SU(4) limit (see Ref.~\onlinecite{marston89} for example) of $v_{F1}=v_{F2}$ and $g_{x\sigma}=g_{t\nu}=g_{xu}=g_{tu2}=0$. Furthermore, previously derived RG equations for half-filled ladders\cite{balents96} with particle-hole symmetry can be recovered by setting $v_{F1}=v_{F2}$.

We integrate Eq.~(\ref{eq:rgflow}) numerically starting from initial values of $g_i(0)$ determined by the bare interactions as presented in the Appendix. In general, the couplings diverge at some large length scale. We integrate Eq.~(\ref{eq:rgflow}) until the largest coupling constant equals 0.01, so that the one loop RG flow equations remain valid. We have checked that the phase diagram so obtained is not sensitive to the choice of infrared cut-off.


\section{Bosonization} \label{sec:bosonization}

To elucidate nature of the phase, we use the Abelian bosonization technique to interpret the action semiclassically. The fermionic field operator $\psi^{}_{nP\sigma}$ is first expressed in terms of dual Hermitian bosonic fields $\phi_{n\sigma}$ and $\theta_{n\sigma}$ by
\begin{equation}\label{eq:field}
\psi_{nP\sigma} 
= \frac{1}{\sqrt{2\pi\epsilon}} \kappa_{n\sigma} 
\exp[i(P\phi_{n\sigma}+\theta_{n\sigma})],
\end{equation}
where $\epsilon$ is a short-distance cutoff, and again $P=R$ or $L=\pm 1$ is chirality.  The bosonic fields satisfy the usual commutation relations
{\setlength\arraycolsep{2pt}
\begin{eqnarray}
&& [ \phi_{n\sigma}(x), \phi_{n'\sigma'}(x') ] 
= [ \theta_{n\sigma}(x), \theta_{n'\sigma'}(x') ] = 0, \nonumber \\
&& [ \phi_{n\sigma}(x), \theta_{n'\sigma'}(x') ] 
= i \pi \delta_{n,n'} \delta_{\sigma,\sigma'} \Theta(x-x'),
\label{eq:commb}
\end{eqnarray}}
where $\Theta(x)$ is the Heaviside function.  The Klein factor $\kappa_{n\sigma}$ is introduced to ensure the correct anticommutation relations of the original fermionic fields; $\kappa_{n\sigma}$ commutes with the bosonic fields, and satisfies
\begin{equation}
\{ \kappa_{n\sigma}, \kappa_{n'\sigma'} \} 
= 2 \delta_{n, n'} \delta_{\sigma, \sigma'}.
\end{equation}

A canonical transformation separates bosons into charge and spin pieces
{\setlength\arraycolsep{2pt}
\begin{eqnarray}
(\phi, \theta)_{n\rho} &=& \frac{1}{\sqrt{2}}
[(\phi, \theta)_{n\uparrow}+(\phi, \theta)_{n\downarrow}],
\nonumber \\
(\phi, \theta)_{n\sigma} &=& \frac{1}{\sqrt{2}}
[(\phi, \theta)_{n\uparrow}-(\phi, \theta)_{n\downarrow}].
\end{eqnarray}}
We also define $(\phi,\theta)_{r\nu}$ as the following:
\begin{equation}
(\phi,\theta)_{r\nu} 
= \frac{1}{\sqrt{2}} 
[ r (\phi, \theta)_{1\nu}+(\phi, \theta)_{2\nu}],
\end{equation}
where  $r$ is either $+$ or $-$. With these definitions, the noninteracting part of the Hamiltonian density has the following form:
{\setlength\arraycolsep{2pt}
\begin{eqnarray}
&&{\mathcal H}_{0}
= \frac{v_{F1}+v_{F2}}{2\pi} \sum_{r,\nu}
[ (\partial_x \phi_{r\nu})^2 + (\partial_x \theta_{r\nu})^2 ] \nonumber \\
&&- \frac{v_{F1} - v_{F2}}{2\pi} \sum_{\nu}
[ (\partial_x \phi_{+\nu})(\partial_x \phi_{-\nu})
+ (\partial_x \theta_{+\nu})(\partial_x \theta_{-\nu}) ]. \nonumber \\
&&
\end{eqnarray}}
The momentum-conserving part of the interaction can be written as the sum of two terms, ${\mathcal H}^{(1)}_{I} = {\mathcal H}^{(1a)}_{I} + {\mathcal H}^{(1b)}_{I}$, where
\begin{equation}
{\mathcal H}^{(1a)}_I 
= \frac{1}{2\pi^2} 
\sum_{r,\nu} A_{r\nu}
[ (\partial_x \phi_{r\nu})^2 - (\partial_x \theta_{r\nu})^2 ]
\end{equation}
with
{\setlength\arraycolsep{2pt}
\begin{eqnarray}
A_{r\rho} &=& - \frac{1}{2} [ g_{1\rho} + g_{2\rho} + 2r g_{x\rho} ],
\nonumber \\
A_{r\sigma} &=& - \frac{1}{8} [ g_{1\sigma} + g_{2\sigma} + 2r g_{x\sigma} ].
\nonumber
\end{eqnarray}}
And
{\setlength\arraycolsep{2pt}
\begin{eqnarray}
{\mathcal H}^{(1b)}_I 
&=& - \frac{1}{(2\pi\epsilon)^2} 
[ 2 \hat{\Gamma} g_{t\sigma} \cos 2 \theta_{-\rho} \cos 2 \phi_{+\sigma}
- (g_{1\sigma} + g_{2\sigma})
\nonumber \\
&& \times \cos 2 \phi_{+\sigma} \cos 2 \phi_{-\rho}
- 2 \hat{\Gamma} g_{x\sigma} \cos 2 \phi_{+\sigma} \cos 2 \theta_{-\sigma}
\nonumber \\
&&+ \cos 2 \theta_{-\rho} 
( \hat{\Gamma} g^{+}_{t} \cos 2 \phi_{-\sigma} 
+ g^{-}_{t} \cos 2 \theta_{-\sigma} ) ],
\end{eqnarray}}
with $g^{\pm}_{t} = g_{t\sigma} \mp 4 g_{t\rho}$ and $\hat{\Gamma} = \kappa_{1\uparrow} \kappa_{1\downarrow} \kappa_{2\uparrow} \kappa_{2\downarrow}$. Finally, the bosonized form of the Umklapp interaction density reads
{\setlength\arraycolsep{2pt}
\begin{eqnarray}
{\mathcal H}^{(3)}_{I}
&=& \frac{2}{(2\pi\epsilon)^2} 
\cos 2 \phi_{+\rho} 
[ 2 \hat{\Gamma} g_{xu} \cos 2 \theta_{-\rho}
+ 2 (g_{tu1}
\nonumber \\
&&+ g_{tu2}) \cos 2 \phi_{+\sigma}
+  g^{+}_{u} \cos 2 \phi_{-\sigma}
+ \hat{\Gamma}  g^{-}_{u} \cos 2 \theta_{-\sigma} ],
\nonumber \\
&&
\end{eqnarray}}
with $g^{\pm}_{u} = (g_{tu1}+g_{tu2}) \pm 4 (g_{tu1}-g_{tu2})$.

Note that since the Hermitian operator $\hat{\Gamma}$ obeys $\hat{\Gamma}^2 = I$, $\hat{\Gamma}$ has eigenvalues $\Gamma = \pm 1$. Furthermore, since $[H,\hat{\Gamma}]=0$, $H$ and $\hat{\Gamma}$ can be simultaneously diagonalized. We choose $\Gamma = 1$ in this paper. See Refs.~\onlinecite{fjaerestad02,tsuchiizu02,wu03} for details regarding to the subtleties of the Klein factors. For the sake of simplicity we also suppress the Klein factors in what follows.

At half-filling the system is either fully gapped (by far the most common case) or gapless in all sectors (referred to as C2S2 -- see Ref.~\onlinecite{balents96}.) We do not find any partially gapped phases at half-filling. For the gapped phases the ground state configuration of the bosonic fields can be determined by minimizing the energy of the low-energy effective Hamiltonian at the end of the RG flow.

\begin{table}
\begin{ruledtabular}
\begin{tabular}{cccccc}
& $\phi_{+\rho}$ & $\phi_{+\sigma}$ & $\theta_{-\rho} $ &
$\theta_{-\sigma}$ & $\phi_{-\sigma}$ \\
\hline
CDW & 0 & 0 & $\pi/2$ & 0 & * \\
SP & $\pi/2$ & 0 & $\pi/2$ & 0 & * \\
SF & 0 & 0 & 0 & 0 & * \\
DC & $\pi/2$ & 0 & 0 & 0 & * \\
D-Mott & 0 & 0 & 0 & * & 0 \\
D$^{\prime}$-Mott & $\pi/2$ & 0 & 0 & * & 0 \\
S-Mott & 0 & 0 & $\pi/2$ & * & 0 \\
S$^{\prime}$-Mott & $\pi/2$ & 0 & $\pi/2$ & * & 0 \\
\end{tabular}
\end{ruledtabular}
\caption{Potential phases of the Cu-O two-leg ladder, and their pinned configurations (a `*' means that the field is fluctuating.)}
\label{table:fieldconfig}
\end{table}

Once the ground state configuration of the bosonic fields is determined, the bosonized order parameters may be examined to determine the physical nature of the ground states.  We consider the same order parameters studied in previous work on one-band ladder systems: the order parameter of the $(\pi,\pi)$ charge density wave (CDW) phase, the $(\pi,\pi)$ spin Peierls (SP) phase, the SF phase, the diagonal current (DC) phase, and the four quantum disordered phases (D-Mott, D$^{\prime}$-Mott, S-Mott, and S$^{\prime}$-Mott.) Order parameters expressed in terms of bosonic fields are as follows:
\begin{subequations}
\label{eq:orderparam}
{\setlength\arraycolsep{2pt}
\begin{eqnarray}
{\mathcal O}_\mathrm{CDW} 
& \propto & 
\cos\phi_{+\rho} \sin\theta_{-\rho} 
\cos\phi_{+\sigma} \cos\theta_{-\sigma} \nonumber \\
& & + \sin\phi_{+\rho} \cos\theta_{-\rho} 
\sin\phi_{+\sigma} \sin\theta_{-\sigma}, \label{eq:ocdw}
\end{eqnarray}}
{\setlength\arraycolsep{2pt}
\begin{eqnarray}
{\mathcal O}_\mathrm{SP} 
& \propto & 
\cos\phi_{+\rho} \cos\theta_{-\rho} 
\sin\phi_{+\sigma} \sin\theta_{-\sigma} \nonumber \\
& & + \sin\phi_{+\rho} \sin\theta_{-\rho} 
\cos\phi_{+\sigma} \cos\theta_{-\sigma}, \label{eq:osp}
\end{eqnarray}}
{\setlength\arraycolsep{2pt}
\begin{eqnarray}
{\mathcal O}_\mathrm{SF} 
& \propto & 
\cos\phi_{+\rho} \cos\theta_{-\rho} 
\cos\phi_{+\sigma} \cos\theta_{-\sigma} \nonumber \\
& & + \sin\phi_{+\rho} \sin\theta_{-\rho} 
\sin\phi_{+\sigma} \sin\theta_{-\sigma}, \label{eq:osf}
\end{eqnarray}}
{\setlength\arraycolsep{2pt}
\begin{eqnarray}
{\mathcal O}_\mathrm{DC} 
& \propto & 
\cos\phi_{+\rho} \sin\theta_{-\rho} 
\sin\phi_{+\sigma} \sin\theta_{-\sigma} \nonumber \\
& & + \sin\phi_{+\rho} \cos\theta_{-\rho} 
\cos\phi_{+\sigma} \cos\theta_{-\sigma}, \label{eq:odc}
\end{eqnarray}}
{\setlength\arraycolsep{2pt}
\begin{eqnarray}
{\mathcal O}_\mathrm{D-Mott} 
& \propto & 
e^{i\theta_{+\rho}} \cos\theta_{-\rho} 
\cos\phi_{+\sigma} \cos\phi_{-\sigma} \nonumber \\
& & -i e^{i\theta_{+\rho}} \sin\theta_{-\rho} 
\sin\phi_{+\sigma} \sin\phi_{+\sigma}, \label{eq:odmott}
\end{eqnarray}}
{\setlength\arraycolsep{2pt}
\begin{eqnarray}
{\mathcal O}_\mathrm{S-Mott} 
& \propto & 
e^{i\theta_{+\rho}} \cos\theta_{-\rho} 
\sin\phi_{+\sigma} \sin\phi_{-\sigma} \nonumber \\
& & -i e^{i\theta_{+\rho}} \sin\theta_{-\rho} 
\cos\phi_{+\sigma} \cos\phi_{-\sigma}. \label{eq:sdmott}
\end{eqnarray}}
\end{subequations}

We emphasize that Klein factors have been taken into account in the derivation of these expressions, though we suppress the Klein factors for the sake of simplicity (see Refs.~\onlinecite{fjaerestad02,tsuchiizu02,wu03} for more details.) Note that ${\mathcal O}_\mathrm{D-Mott}$ is the order parameter for both D-Mott and D$^{\prime}$-Mott phase (the same holds between S-Mott and S$^{\prime}$-Mott.) These primed phases are half-cell translated states of the unprimed states since $\phi_{+\rho}$ differs by $\pi/2$.\cite{tsuchiizu02} The bosonized form of the order parameters is essentially the same as in one band case except for the appearance of complicated band coefficients $a_{ij}$. 

Table~\ref{table:fieldconfig} lists the bosonic configurations for all the gapped phases. Note that since $\phi_{r\nu}$ and $\theta_{r\nu}$ are conjugate to each other, they cannot be pinned simultaneously. The nature of these phases was discussed in Refs.~\onlinecite{lin98,fjaerestad02,tsuchiizu02,wu03} in one band context. Transitions between the various phases have the same critical behavior as in the one-band case (see, for example, Fig.~5 in Ref.~\onlinecite{tsuchiizu02}.)

Note also that the order parameter of the circulating current phase that preserves translational symmetry does not appear in Eqs.~(\ref{eq:orderparam}). This is because the long-wavelength part of the current is determined by gradients of the boson fields, $\partial_x \theta_{\pm \rho}$, but these cannot acquire non-zero expectation values at the semiclassical minima. Field $\theta_{-\rho}$ locks into the particular values specified in Table \ref{table:fieldconfig}, and field $\theta_{+\rho}$ fluctuates with zero mean gradient. The bosonized form of the current along any link inside the unit cell, apart from the unimportant gradient terms, is given by
{\setlength\arraycolsep{2pt}
\begin{eqnarray}
j_{ij}(x) 
&\propto& 
(-1)^x t_{ij} ( a_{i1} a_{j2} - a_{i2} a_{j1} )
\nonumber \\
&&\times ( \cos\phi_{+\rho} \cos\theta_{-\rho} 
\cos\phi_{+\sigma} \cos\theta_{-\sigma}
\nonumber \\
&& + \sin\phi_{+\rho} \sin\theta_{-\rho} 
\sin\phi_{+\sigma} \sin\theta_{-\sigma} ).
\end{eqnarray}}
It is clear from this expression that the current pattern, if it exists, must be staggered along the leg direction such that it breaks translational symmetry. Hence, the circulating current phase cannot occur at weak coupling in half-filled Cu-O two-leg ladder system. Furthermore, this is also true away from half-filling, at least as long as the Fermi level intersects only two bands.


\section{Weak Coupling Phase Diagrams} \label{sec:weak}

\begin{table}
\begin{ruledtabular}
\begin{tabular}{ccccc}
& D-Mott & D$^{\prime}$-Mott & S-Mott & S$^{\prime}$-Mott \\
\hline
Volume(\%) & 40.78 & 3.96 & 12.7 & 22.97 \\
\hline \hline
& SF & SP & CDW & C2S2 \\
\hline
Volume(\%) & 0.84 & 5.63 & 12.86 & 0.19\\
\end{tabular}
\end{ruledtabular}
\caption{Volume of each phase over the entire parameter space determined by Monte-Carlo sampling in the five orbital case. We sample $U_d$, $U_p$, $V^{\parallel}_{dp}$, $V^{\perp}_{dp}$, and $V^{}_{pp}$ uniformly.}
\label{table:phasevol1}
\end{table}

Equipped with the above considerations, we deduce the phase diagram of the three-band Cu-O ladder system in the weak coupling regime. By sampling the parameter space at random, we may quantify the volume of each phase. Table~\ref{table:phasevol1} lists the volume in the five orbital case. We find that the D-Mott, S$^{\prime}$-Mott, S-Mott, and CDW phases occupy most of the parameter space. The SF phase also arises, though its volume is very small. However, the DC phase does not appear.

\begin{figure}
\includegraphics[width=3.0in]{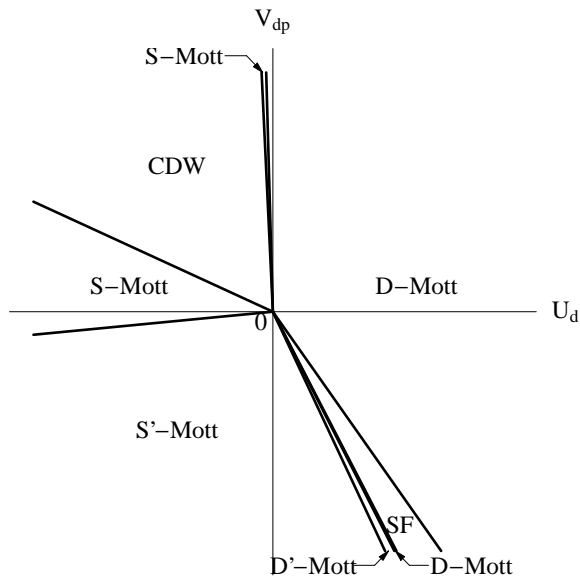}
\caption{Weak coupling phase diagram in the five orbital case: $U_p = 0.38 U_d$ and $V_{pp} = 0.1 V_{dp}$.}
\label{fig:wphase1}
\end{figure}

It is also useful to consider some slices through parameter space that may be relevant for real systems. In Fig.~\ref{fig:wphase1} we present the weak coupling phase diagram of the five orbital Cu-O ladder for $U_p = 0.38 U_d$ and $V_{pp} = 0.1 V_{dp}$. In the absence of nearest neighbor Coulomb interactions the D-Mott phase appears for positive on-site Coulomb interactions; the S-Mott phase appears for negative on-site Coulomb interactions.\cite{withoutv} The CDW phase is found in the second quadrant bisecting the S-Mott phase. There is a small region of S-Mott phase located around $92\degree$ between the D-Mott phase and the CDW phase. Note that direct transitions between the D-Mott and the CDW are not generically possible since it is necessary to unpin two bosonic field simultaneously: $\theta_{-\rho}$ and $\phi_{-\sigma}$ or $\theta_{-\sigma}$ (In the one-band model, for example, the SF phase mediates the transition between the D-Mott and the CDW phase.\cite{fjaerestad02}). The SF phase arises next to the D-Mott phase in the fourth quadrant bisecting the D-Mott phase. A tiny portion of the D-Mott phase exists between the D$^{\prime}$-Mott and the SF phases, which can be understood with the same argument used for the transition between the D-Mott and the CDW: a direct transition from the S$^{\prime}$-Mott to the D-Mott or the SF phase is not possible since two bosonic fields must become unpinned simultaneously. Increasing $V_{pp}$ and holding other parameters fixed, the SF phase shrinks and then vanishes. Except for the disappearance of the SF phase, the phase diagram is qualitatively similar to Fig.~\ref{fig:wphase1} with slightly different phase boundaries.

\begin{figure}
\includegraphics[width=3.0in]{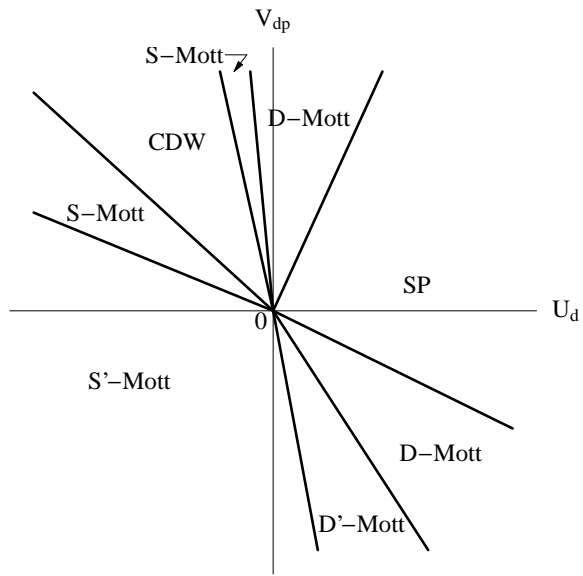}
\caption{Weak coupling phase diagram in the seven orbital case: $U_p = 0.38 U_d$ and $V_{pp} = 0.1 V_{dp}$.}
\label{fig:wphase2}
\end{figure}

The phase diagram of the seven orbital case is presented in Fig.~\ref{fig:wphase2}. The phase diagram is topologically rather similar to the five orbital case, however, with two notable differences: the SP phase now appears inside the D-Mott phase, and the SF phase disappears altogether. Both of these aspects might be related to the anisotropy between the effective exchange integrals parallel and perpendicular to the legs. In a DMRG calculation on a two-leg Cu-O ladder system by Nishimoto {\em et al.},\cite{nishimoto02} it was found that in the absence of the outer O orbitals, the dimensionless anisotropy ratio $R = \langle {\bm S}_{i} \cdot {\bm S}_{j} \rangle_{\mathrm{rung}} / \langle {\bm S}_{i} \cdot {\bm S}_{j} \rangle_{\mathrm{leg}}$ becomes significantly larger than one. Nishimoto {\it et al.} concluded that the inclusion of outer oxygens is crucial for a qualitatively correct description of two-leg Cu-O ladders. We can understand this behavior as follows. When $R$ is smaller and the system is therefore more isotropic, the SP phase seems to be a more stable state than the D-Mott phase, as the D-Mott phase may be thought of as a product of rung singlets. For large values of the anisotropy ratio, the strong effective Heisenberg interaction along the rung makes rung singlets relatively strong, favoring the D-Mott phase.


\section{Strong Coupling and the Role of the Exchange Energies}
\label{sec:strong}

\begin{figure}
\includegraphics[width=3.0in]{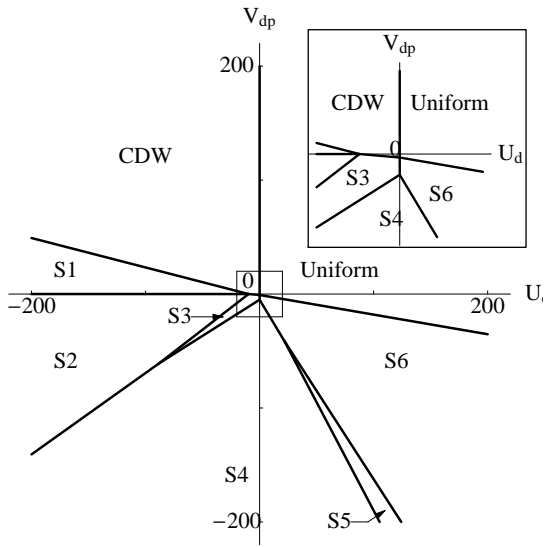}\\
\caption{Strong coupling phase diagram of the five orbital case:
Parameter values are $\epsilon = 3.0$ eV, $U_p = 0.38 U_d$, and $V_{pp} = 0.1 V_{dp}$. Inset is an enlarged plot which is drawn from -20 to 20 eV. Phase $S1$ through $S6$ are defined in the text.}
\label{fig:strongphase1}
\end{figure}

In the strong coupling, atomic, limit of the extended Hubbard model, hopping matrix elements may be ignored, and the system is purely electrostatic. The number of electrons on each orbital is restricted to be an integer 0, 1, or 2 by the Pauli exclusion principle. We assume periodic boundary conditions and take the number of sites to be large. The various charge-ordered ground states of the Cu-O ladder may then be found by energy minimization. The lowest energy charge configurations at most double the size of the unit cell in size. Results for the five orbital case are shown in Fig.~\ref{fig:strongphase1}.

For positive $U$ and $V$ we find a uniform phase in which each Cu site is occupied by one electron. This phase corresponds to a Mott insulator. Superexchange Heisenberg spin-spin interaction $J^\parallel$ and $J^\perp$ between nearest-neighbor copper spins are induced, respectively, along the legs and rungs of the ladder perturbatively at fourth order in the hopping.\cite{exchange} In the five orbital case
{\setlength\arraycolsep{2pt} 
\begin{eqnarray}
J^{\parallel}
&=& \frac{4 t_{dp}^4}{(\epsilon+U_d-U_p+3V_{dp}-4V_{pp})^2}
\nonumber \\
&&\times \left(\frac{1}{U_d} 
+ \frac{2}{2\epsilon+2U_d-U_p+4V_{dp}-8V_{pp}} \right),
\nonumber \\
&& \\
J^{\perp}
&=&
\frac{4 t_{dp}^4}{(\epsilon+U_d-U_p+3V_{dp}-8V_{pp})^2}
\nonumber \\
&&\times \left(\frac{1}{U_d} 
+ \frac{2}{2\epsilon+2U_d-U_p+4V_{dp}-16V_{pp}} \right),
\nonumber
\end{eqnarray}}
and in the seven orbital case
{\setlength\arraycolsep{2pt}
\begin{eqnarray}
J^{\parallel}
&=& J^{\perp}
= \frac{4 t_{dp}^4}{(\epsilon+U_d-U_p+7V_{dp}-8V_{pp})^2}
\nonumber \\
&&\times \left( \frac{1}{U_d} + 
\frac{2}{2\epsilon+2U_d-U_p+8V_{dp}-16V_{pp}} \right),
\nonumber \\
&&
\end{eqnarray}}
where the on-site energy difference between the $d$ and $p$ electrons is $\epsilon = \epsilon_d - \epsilon_p > 0$. With these effective superexchange interactions, degeneracies in the spin degree of freedom are lifted with antiferromagnetic interactions appearing when $U_d > U_p > 0$ and $V_{dp} > V_{pp} > 0$. It is easy to see that the anisotropy ratio is larger than one in five orbital case within this simple calculation, qualitatively consistent with the previous DMRG calculation.\cite{nishimoto02} This phase can then be mapped to the D-Mott phase found at weak coupling in a similar region of parameter space. In the seven orbital case, the exchanges are isotropic, weakening tendencies towards a D-Mott phase. It should be noted, however, that if we set $V_{pp}=0$ in the five orbital case, the system becomes isotropic $J^{\parallel}=J^{\perp}$ but the SP phase still does not arise. Another possible way to explain the SP phase is to consider frustration. For example, competition between spin exchange along the plaquette and along the diagonal can stablize the SP phase. Also, as pointed out by Kivelson {\em et al.}\cite{kivelson04} other higher order interactions such as plaquette four-spin ring exchange interaction can be comparable in magnitude to the superexchange energy even at relatively large $U$, and other phases may arise due to the competition between superexchange and ring exchange interactions.

For $U < 0$, the $(\pi,\pi)$ CDW phase arises in a similar region as that found in the case of weak coupling. It is robust against perturbation due to the hopping term. We also find other charge ordered phases that have no direct relationship to the weak coupling phase diagram. All of these charge ordered phases have tendencies toward stripes or charge separation. Phase $S1$, near the negative $U$ axis with positive $V$, has two holes occupying the O $p_{y}$ orbital. The ladder breaks into two decoupled completely filled chains. In the $S2$ regime, two holes occupy O $p_{x}$ orbitals on a leg in doubled unit cell. In the $S3$ regime, pairs of two holes occupy the Cu orbital and the O $p_{x}$ orbitals on a leg in doubled unit cell. In $S4$ and $S5$, holes gather around a Cu site. Finally, in the $S6$ regime holes gather along one leg of the ladder, and the other leg is completely filled.


\section{Conclusion} \label{sec:conclusion}

We determined the phase diagram of the half-filled Cu-O two-leg ladder both at weak and strong coupling. At weak coupling, perturbative RG flows are interpreted with the use of bosonization. Due to absence of particle-hole symmetry the Fermi velocities of the two bands differ to each other and different RG flow equations govern the three band model than in the case of the simple one-band ladder. After bosonization, however, the Hamiltonian and the order parameters have essentially the same form as in the one-band case; it is the phase diagrams that differ qualitatively. By studying the Cu-O ladder with and without outer oxygen sites, four interesting conclusions may be drawn. First, in the physically relevant regime, the D-Mott phase arises when outer oxygen sites are absent. Second, the SP phase appears in the seven orbital case. Third, the SF phase only appears when the outer oxygen sites are absent. The appearance / disappearance of phases may be related to the anisotropy of the effective exchange energies in the system. Finally, we found that the CC phase which preserves lattice translational symmetry does not appear, at least at weak coupling in the Cu-O two-leg ladder. Local currents or charge modulation, when they arise, always have a staggered pattern.


\section{Acknowledgments}

We would like to thank M. Tsuchiizu for pointing out the existence of the S-Mott phase between the D-Mott and CDW phases and C. M. Varma for a helpful discussion. This work was supported in part by National Science Foundation Grant No. DMR-0213818 (S.L. and J.B.M.), the Galkin Foundation (S.L) and funds from the David Saxon chair at UCLA (J.O.F.) J.O.F. also thanks the Australian Research Council for financial support, and the Rudolf Peierls Centre for Theoretical Physics at Oxford University for its hospitality.


\appendix*
\section{Relation between the continuum couplings and the lattice model interaction parameters}

In this appendix we discuss the relations between the continuum coupling constants $\lambda$ and $g$ and the lattice Coulomb interactions $U$ and $V$. Define $w^{(0)}_{ij}$ and $w^{(1)}_{ij}$ as follows:
\begin{subequations} \label{eq:w}
\begin{equation} \label{eq:w1}
w^{(0)}_{ij} = U_{i} \delta_{ij} + V^{intra}_{ij},
\end{equation}
\begin{equation} \label{eq:w2}
w^{(1)}_{ij} = V^{inter}_{ij},
\end{equation}
\end{subequations}
where $U_i$, $V^{intra}_{ij}$, and $V^{inter}_{ij}$ are given in Eqs.~(\ref{eq:ui}) and (\ref{eq:nnv}). Parameters $w^{(0)}$ and $w^{(1)}$ represent the intracell and intercell interactions, respectively. Double-counting is avoided by taking $w^{(q)}_{ij}$ to not be symmetric. We introduce coefficients $f(i,j,k,l,m,n,p,q)$ as
{\setlength\arraycolsep{2pt}
\begin{eqnarray}
&&f(i,j,k,l,m,n,p,q) 
\nonumber \\
&&= - a_{ik} a_{il} a_{jm} a_{jn} \cos[(k_{Fm} + p k_{Fn})q] w^{(q)}_{ij},
\end{eqnarray}}
where $i$ and $j$ are site indices, $k$, $l$, $m$, and $n$ are band indices, $p=\pm1$, and $q = 0,1$.

With these definitions, the continuum coupling constants $\tilde{\lambda}$ and $\tilde{g}$ are given by
\begin{displaymath}
\tilde{\lambda}_{1\rho} = \sum_{i,j,q} f(i,j,1,1,1,1,-1,q),
\end{displaymath}
\begin{displaymath}
\tilde{\lambda}_{2\rho} = \sum_{i,j,q} f(i,j,2,2,2,2,-1,q),
\end{displaymath}
{\setlength\arraycolsep{2pt}
\begin{eqnarray}
\tilde{\lambda}_{x\rho}
&=& \sum_{i,j,q} f(i,j,1,1,2,2,-1,q) + f(i,j,2,2,1,1,-1,q)
\nonumber \\
& & - f(i,j,1,2,2,1,-1,q),
\nonumber
\end{eqnarray}}
\begin{displaymath}
\tilde{\lambda}_{1\sigma} = \tilde{\lambda}_{2\sigma} = 0,
\end{displaymath}
\begin{displaymath}
\tilde{\lambda}_{x\sigma} = -4 \sum_{i,j,q} f(i,j,1,2,2,1,-1,q),
\end{displaymath}
\begin{displaymath}
\tilde{g}_{1\rho} = \sum_{i,j,q} 2 f(i,j,1,1,1,1,-1,q) - f(i,j,1,1,1,1,1,q),
\end{displaymath}
\begin{displaymath}
\tilde{g}_{2\rho} = \sum_{i,j,q} 2 f(i,j,2,2,2,2,-1,q) - f(i,j,2,2,2,2,1,q),
\end{displaymath}
{\setlength\arraycolsep{2pt}
\begin{eqnarray}
\tilde{g}_{x\rho}
&=& \sum_{i,j,q} f(i,j,1,1,2,2,-1,q) + f(i,j,2,2,1,1,-1,q)
\nonumber \\
&& - f(i,j,1,2,2,1,1,q),
\nonumber
\end{eqnarray}}
\begin{displaymath}
\tilde{g}_{1\sigma} = -4 \sum_{i,j,q} f(i,j,1,1,1,1,1,q),
\end{displaymath}
\begin{displaymath}
\tilde{g}_{2\sigma} = -4 \sum_{i,j,q} f(i,j,2,2,2,2,1,q),
\end{displaymath}
\begin{displaymath}
\tilde{g}_{x\sigma} = -4 \sum_{i,j,q} f(i,j,1,2,2,1,1,q),
\end{displaymath}
\begin{displaymath}
\tilde{g}_{t\rho} = \sum_{i,j,q} 2 f(i,j,1,2,1,2,-1,q) - f(i,j,1,2,1,2,1,q),
\end{displaymath}
\begin{displaymath}
\tilde{g}_{t\sigma} = -4 \sum_{i,j,q} f(i,j,1,2,1,2,1,q),
\end{displaymath}
\begin{displaymath}
\tilde{g}_{xu} = \sum_{i,j,q} 2 f(i,j,1,2,1,2,1,q),
\end{displaymath}
\begin{displaymath}
\tilde{g}_{tu1} = \sum_{i,j,q} f(i,j,1,1,2,2,1,q) + f(i,j,2,2,1,1,1,q),
\end{displaymath}
\begin{equation}\label{eq:coupling}
\tilde{g}_{tu2} = -2 \sum_{i,j,q} f(i,j,1,2,2,1,1,q).
\end{equation}
The sum in each of the above equations runs over all site indices in the unit cell. Equations (\ref{eq:coupling}) are the most general form of coupling constants for the case of two bands cutting the Fermi level and short range Coulomb interactions. By modifying $w^{(0)}$ or $w^{(1)}$ further generalizations can be studied.


\begin{thebibliography}{47}
\bibitem{current_algebra} Current operators in this section should not be confused with the physical current operators.
\bibitem{withoutv} This happens for more general cases in which $|U_d| > |U_p|$ and also in the absence of the nearest-neighbor Coulomb interactions.
\bibitem{exchange} Note that these expressions look different than those found in the literature because the particle picture is used here. By performing a particle-hole transformation the more common form of the exchange energies can be recovered.
\end{thebibliography}
\end{document}